\documentclass[nonacm, sigplan]{acmart}
\makeatletter
\def\@ACM@checkaffil{% Only warnings
    \if@ACM@instpresent\else
    \ClassWarningNoLine{\@classname}{No institution present for an affiliation}%
    \fi
    \if@ACM@citypresent\else
    \ClassWarningNoLine{\@classname}{No city present for an affiliation}%
    \fi
    \if@ACM@countrypresent\else
        \ClassWarningNoLine{\@classname}{No country present for an affiliation}%
    \fi
}
\makeatother
\usepackage{graphicx} % what this pacakge is for
\usepackage{wrapfig}
\usepackage{tcolorbox}
\usepackage{subcaption}
\usepackage{enumitem}
\setitemize{noitemsep,topsep=0pt,parsep=0pt,partopsep=0pt}
\setlist[itemize]{leftmargin=*}
\setlist[enumerate]{leftmargin=*}
\AtBeginDocument{%
  }

\usepackage{graphicx} % what this pacakge is for
\usepackage{wrapfig}
\usepackage{tcolorbox}
\usepackage{subcaption}
\usepackage{enumitem}
\setitemize{noitemsep,topsep=0pt,parsep=0pt,partopsep=0pt}
\setlist[itemize]{leftmargin=*}
\setlist[enumerate]{leftmargin=*}
\preto{\abstractkeywords}{\nolinenumbers}

\begin{document}

\title{KiSS: A Novel Container Size-Aware Memory Management Policy for Serverless in Edge-Cloud Continuum}
%%\subtitle{\normalsize{ISCA 2025 Submission
%%    \textbf{\#NaN} -- Confidential Draft -- Do NOT Distribute!!}}
%\author{\normalsize{ISCA 2025 Submission
 %   \textbf{\#NaN} -- Confidential Draft -- Do NOT Distribute!!}}
\author{Sabyasachi Gupta\textsuperscript{\symbol{42}\dag}}
  \affiliation{
  \institution{Texas A\&M University}}
  \email{sabyasachi.gupta@tamu.edu}

\author{Paul Gratz\textsuperscript{\symbol{42}\dag}}
  \affiliation{
  \institution{Texas A\&M University}}
  \email{pgratz@tamu.edu@tamu.edu}

\author{John Lusher\textsuperscript{\symbol{42}\dag}}
  \affiliation{
  \institution{Texas A\&M University}}
  \email{john.lusher@tamu.edu}

\begin{abstract}
Serverless computing has revolutionized cloud architectures by enabling developers to deploy event-driven applications via lightweight, self-contained virtualized containers. However, serverless frameworks face critical cold-start challenges in resource-constrained edge environments, where traditional solutions fall short. The limitations are especially pronounced in edge environments, where heterogeneity and resource constraints exacerbate inefficiencies in resource utilization.

This paper introduces KiSS (Keep it Separated Serverless), a static, container size-aware memory management policy tailored for the edge-cloud continuum. The design of KiSS is informed by a detailed workload analysis that identifies critical patterns in container size, invocation frequency, and memory contention. Guided by these insights, KiSS partitions memory pools into categories for small, frequently invoked containers and larger, resource-intensive ones, ensuring efficient resource utilization while minimizing cold starts and inter-function interference. Using a discrete-event simulator, we evaluate KiSS on edge-cluster environments with real-world-inspired workloads. 

Results show that KiSS reduces cold-start percentages by 60\% and function drops by 56.5\%, achieving significant performance gains in resource-constrained settings. This work underscores the importance of workload-driven design in advancing serverless efficiency at the edge.
\end{abstract}


\maketitle

\section{Introduction}\label{sec:intro}

Serverless computing simplifies deploying event-driven applications by removing infrastructure management. Platforms such as AWS Lambdas~\cite{aws_lambda_run_2024}, Google Cloud Functions~\cite{google_cloud_google_2024}, and Azure Functions~\cite{miscrosoft_ignite_azure_2023} dynamically allocate resources to meet fluctuating demand, making them particularly attractive for workloads that are unpredictable~\cite{ao_faasnet_2021,ustiugov_benchmarking_2021}. These economic efficiencies reduce costs associated with idle resources, driving widespread adoption across industries. Serverless has enabled diverse applications, from large-scale data processing and machine learning to real-time analytics and IoT (Internet of Things). According to Gartner, global cloud spending increased by over 20\% in 2022, highlighting the scalability and flexibility offered by serverless computing and other cloud-native models.

Despite its scalability and flexibility, traditional serverless computing faces critical performance challenges, particularly in high-demand and emerging edge node scenarios. Cold start delays disrupt real-time responsiveness in latency-sensitive applications such as financial transactions and IoT analytics~\cite{alexander_faascache_2021,ustiugov_benchmarking_2021}. To address these issues in typical high-demand, cloud scenarios, solutions such as caching (e.g., FaaSCache, SONIC)~\cite{ashraf_sonic_2021, alexander_faascache_2021} and snapshot-based state restoration (e.g., FaaSnap, Catalyzer)~\cite{ao_faasnap_2022, dong_catalyzer_2020} have been proposed to reduce initialization delays. Techniques like predictive pre-warming and hardware-optimized memory access (e.g., REAP)~\cite{ustiugov_benchmarking_2021} also show promise but often rely on specialized infrastructure, limiting their adaptability to other environments. These challenges, while partially mitigated in cloud settings, become significantly more pronounced in resource-constrained edge environments.

An emerging application space for the FaaS model lies in extending function execution to the edge node for quicker response times and greater data privacy~\cite{rausch_optimized_2021, xie_when_2021}. Edge FaaS, which brings computation closer to data sources, is increasingly adopted for latency-sensitive applications such as industrial IoT, autonomous systems, and smart cities. However, edge environments lack the computational slack of cloud data centers to maintain large pre-warmed pools or deploy infrastructure-heavy optimizations. When it comes to heterogeneous devices, the diversity in hardware capabilities and network conditions across edge nodes complicates function placement and resource scheduling. Unlike cloud setups, edge nodes cannot over-provision resources to accommodate workload surges, instead functions which cannot execute immediately must be punted up to the cloud, obviating the gains of edge FaaS. Thus adaptive scaling in the edge node FaaS is critical.

\begin{figure}[tb!] 
    \centering 
    \includegraphics[trim = 15 350 15 15, width= 1\columnwidth]{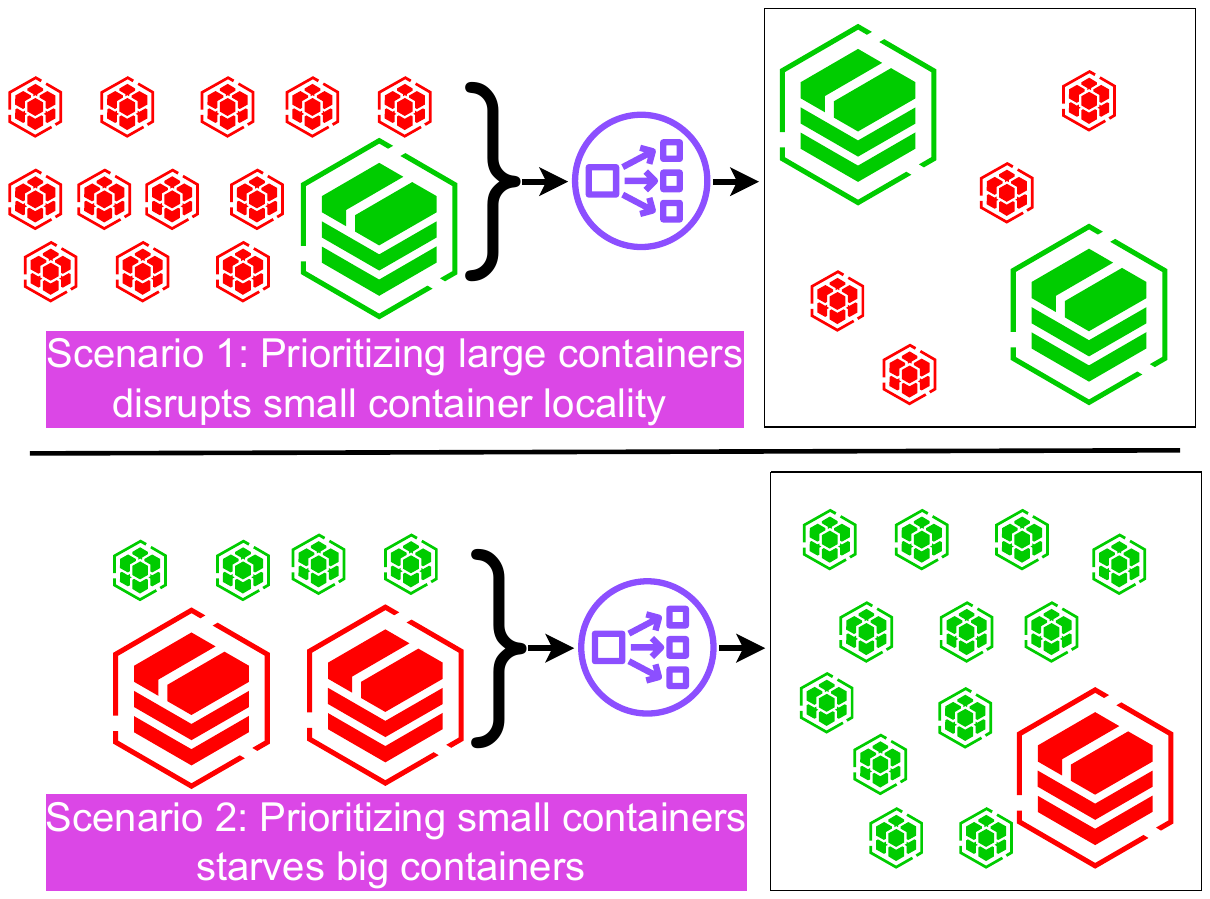} 
    \caption{Memory contention in warm pools: (a) Large containers displace small containers, disrupting locality and increasing cold starts; (b) Small containers dominate due to high invocation ratio starving Large containers.} 
    \label{fig:intro_pools} 
\end{figure}
\subsection{Inter-Function Memory Contention: A Key Issue}
Among the challenges in Edge environments, inter-function memory contention stands out as one of the most significant problems. Warm memory pools, designed to minimize cold starts, are often dominated by small containers due to their higher invocation frequency. This imbalance restricts large containers from entering the warm pool, leading to frequent cold starts. Conversely, when large containers manage to enter the warm pool, they displace multiple small containers, disrupting their temporal locality and resulting in cascading cold starts for frequently invoked small functions.

Furthermore, function chaining, a common pattern in serverless workflows,\textbf{ exacerbates this issue}. Research on chaining frameworks such as Xanadu~\cite{daw_xanadu_2020} and SpecFaaS~\cite{stojkovic_specfaas_2023} highlight the importance of maintaining temporal locality in warm pools to avoid unnecessary cold start penalties. Without effective memory management, the performance of interconnected functions deteriorates, undermining the responsiveness of latency-sensitive applications.

As illustrated in Figure~\ref{fig:intro_pools}, this dynamic leads to cascading inefficiencies. Small containers, which dominate warm pools, exclude larger containers, forcing them into frequent cold starts. When larger containers enter the pool, they displace numerous small containers, degrading their performance by increasing cold start percentages and disrupting their temporal locality. Addressing this imbalance is critical for improving latency-sensitive applications in edge environments. 

In this paper we introduce a new, FaaS memory management policy cognizant of the resource constraints of FaaS in the edge node.  \textit{KiSS} (Keep It Separated Serverless) breaks the available memory into two pools, preventing the interference between large, less frequently used containers and smaller, higher locality containers.  The paper makes the following contributions:
\begin{itemize}
    \item The first work to examine memory management for FaaS in highly resource constrained, edge node systems and identify the interaction between larger and smaller containers in these environments.
    \item Introduces a new, container size aware, memory management policy for edge node FaaS systems.
    \item Shows that KiSS reduces cold start latency by up-to 60\% while also reducing the number of times functions must be dropped (pushed to the cloud for execution) by up-to 56.5\%.
\end{itemize}

\section{Background}\label{sec:backgrnd}

\subsection{Cold Start Latency and Mitigation Techniques}

Traditional FaaS platforms mitigate cold starts through snapshotting, lightweight virtualization, and warm-state management. Snapshot-based methods like \textbf{REAP} and \textbf{Catalyzer} reduce initialization time by preloading or restoring container states but require significant memory and I/O resources, limiting scalability~\cite{dong_catalyzer_2020, ustiugov_benchmarking_2021}. Lightweight virtualization solutions, such as \textbf{Firecracker} microVMs, achieve fast startup times with strong isolation but depend on robust infrastructure, making them less adaptable to fluctuating workloads~\cite{agache_firecracker_2020}. Warm-state management techniques like \textbf{Faa\$T}~\cite{romero_faa_2021} and \textbf{Kraken}~\cite{vivek_kraken_2021} keep frequently invoked containers ready, balancing readiness and cost efficiency under predictable workloads but incurring overhead when demand is erratic~\cite{romero_faa_2021, vivek_kraken_2021}. While these methods perform well in resource-rich cloud environments, their resource intensity challenges applicability in edge settings.

\subsubsection{Edge FaaS Perspective}

In edge environments, cold start mitigation emphasizes lightweight designs, resource sharing, and hybrid task distribution. Lightweight execution environments like unikernels~\cite{edward_sock_2018} and \textbf{Firecracker}~\cite{agache_firecracker_2020}, as used by \textbf{TinyFaaS}~\cite{pfandzelter_tinyfaas_2020}, minimize resource usage and initialization delays but require careful orchestration to avoid resource contention. Function co-location, demonstrated by \textbf{Photons}~\cite{v_dukic_photons_2020}, reduces redundant initializations by sharing runtime resources among related functions, though this complicates isolation in multi-tenant setups~\cite{v_dukic_photons_2020}. Hybrid offloading frameworks like \textbf{GeoFaaS}~\cite{malekabbasi_geofaas_2024} balance edge-cloud workloads by offloading latency-tolerant tasks to the cloud and reserving edge resources for real-time operations, requiring reliable connectivity and efficient task management. These edge-specific strategies address cold starts effectively but introduce challenges in scalability and orchestration.

\subsection{Predictive Scaling and Caching Techniques}

Efficient resource allocation is vital for maintaining low latency and high availability in serverless platforms. Predictive scaling and caching techniques dynamically provision resources and reduce cold start latency by leveraging workload prediction and state retention.
Traditional FaaS platforms use predictive scaling and caching to optimize resources, employing techniques (OFC, FaasCache) to reduce cold starts. However, these methods rely on centralized orchestration and workload predictability, limiting their effectiveness in dynamic, resource-constrained edge environments.

\subsubsection{Edge FaaS Perspective}

Edge FaaS platforms adapt predictive scaling and caching techniques to constrain resources and heterogeneous environments. \textbf{EDGE-Cache}~\cite{kim_delay-aware_2022} uses traffic profiling to selectively retain high-priority functions, reducing memory overhead while maintaining readiness for frequent requests. Hybrid frameworks like \textbf{GeoFaaS}~\cite{malekabbasi_geofaas_2024} implement distributed caching to balance resources between edge and cloud nodes, enabling low-latency processing for critical tasks while offloading less critical workloads. Machine learning methods, such as clustering-based workload predictors~\cite{gao_machine_2020} and GRU-based models~\cite{guo_applying_2018}, enhance resource provisioning in edge systems by efficiently forecasting workload spikes. These innovations effectively address cold start challenges in edge environments, though their dependency on accurate predictions and robust orchestration poses scalability challenges.

\subsection{Decentralized Orchestration, Function Placement, and Scheduling}

Efficient orchestration in serverless platforms involves workload distribution, resource optimization, and performance assurance. While traditional FaaS platforms rely on centralized control, edge environments require decentralized and adaptive strategies to address unique challenges such as resource constraints and heterogeneous hardware.

\subsubsection{Edge FaaS Perspective}

Edge FaaS platforms adopt decentralized and adaptive orchestration frameworks to meet the demands of resource-constrained environments. Systems like \textbf{Wukong} distribute scheduling across edge nodes, enhancing data locality and scalability while reducing network latency. Lightweight frameworks such as \textbf{OpenWhisk Lite}~\cite{kravchenko_kpavelopenwhisk-light_2024} optimize resource allocation by decentralizing scheduling policies, minimizing cold starts and latency in edge setups~\cite{benjamin_wukong_2020}. Hybrid solutions like \textbf{OpenFaaS}~\cite{noauthor_openfaasfaas_2024} and \textbf{EdgeMatrix}~\cite{shen_edgematrix_2023} combine edge-cloud orchestration to balance resource utilization, retaining latency-sensitive functions at the edge while offloading non-critical workloads to the cloud. While these approaches improve flexibility, they face challenges in maintaining coordination and ensuring consistent performance across distributed nodes.

\subsection{Challenges \& Research Gaps in Edge Serverless Computing}

The above sections have discussed significant advancements in mitigating cold start latency, scaling, and orchestration for serverless platforms. But since these impose unique challenges to edge environments (limited resources, workload  variability, and stringent latency requirements). Addressing these gaps requires a deeper understanding of the limitations of current solutions and their implications for edge Function-as-a-Service (FaaS)~\cite{rausch_towards_2019}.

\subsubsection{Cold Start Latency in Edge Environments}

As higlighted, cold starts remain a major challenge in edge systems due to limited computational slack. Unlike cloud platforms, which can afford to allocate idle resources, edge devices operate under strict constraints, worsening pre-warming misses during unpredictable traffic. For example, snapshotting techniques like \textbf{FaaSnap}~\cite{ao_faasnap_2022} improve initialization but require considerable memory resources, making them less suitable for constrained environments. The absence of lightweight and adaptive mechanisms for maintaining warm states leads to increased latencies, particularly for latency-critical tasks like IoT or real-time analytics.

\subsubsection{Inefficient Resource Utilization and Placement}

Heterogeneous edge nodes complicate resource allocation and function placement, often leading to contention or under utilization. Current efforts predominantly focus on deriving optimal placement strategies, yet these approaches are computationally expensive and often impractical due to the NP-hard nature of the problem. Therefore, the focus must shift toward lightweight frameworks which can prioritize efficient resource usage, increasing the capacity of edge nodes to handle more requests.

\subsubsection{Workload Variability and Adaptability}

Edge workloads differ fundamentally from cloud workloads in terms of scope and size. Cloud-optimized policies typically adapt to traffic variability over longer invocation periods, which is infeasible for edge environments where workloads change rapidly and unpredictably. Studies such as those on \textbf{Hermod}~\cite{kostis_hermod_2022} reveal that load- and locality-aware schedulers outperform static policies by dynamically adjusting to real-world workload patterns. However, these approaches require further refinement to minimize the overhead introduced by frequent scaling adjustments. Without effective adaptation, edge systems risk over-provisioning during low traffic and queuing delays during demand surges.

\subsubsection{Platform Agnostic Orchestration for Edge}

Platform-agnostic orchestration frameworks, which enable uniform deployment across diverse devices and architectures, are essential for commercial and market adaptation of edge solutions. Existing systems often rely on proprietary APIs or customized platforms, reducing their portability and hindering widespread adoption. For example, lightweight FaaS platforms like \textbf{TinyFaaS}~\cite{pfandzelter_tinyfaas_2020} demonstrate the feasibility of platform-agnostic designs by focusing on resource efficiency and modularity. Such frameworks not only improve scalability but also lower the entry barrier for integrating edge FaaS solutions into diverse environments, fostering greater innovation and adoption.

\subsection{Workload Analysis}\label{sec:work_anly}
By analyzing workload traces and identifying trends, we can glean valuable insights into invocation patterns, resource usage, and scaling inefficiencies. This understanding will inform the design of adaptive and lightweight edge FaaS policies that optimize resource utilization, minimize cold starts, and enhance platform portability. 
Function-as-a-Service (FaaS) workload consist of applications with different sized containers.
Small, lightweight and stateless containers are invoked more frequently than larger ones. Their high frequency requires persistent caching to reduce cold starts and ensure low latency for critical applications. 
On the other hand, large containers have significant memory and dependency requirements but exhibit infrequent invocations. Their cold starts impose longer delays, which require demand-driven strategies to balance responsiveness and resource efficiency.

We conduct a workload analysis to gain a deeper understanding of Function-as-a-Service (FaaS) workload dynamics. Using the Azure functions dataset (2019)~\cite{mohammad_serverless_2020} available for 2 weeks, we identify critical function characteristics—\textit{memory footprint}, \textit{invocation frequency} along with patterns, and \textit{execution time}—as foundational dimensions for efficient resource orchestration. These insights form the basis of the KiSS framework's multi-level warm pool design, enabling it to address the gaps, specifically for edge environments.

\subsubsection{Memory Footprint}
Existing solutions in the literature fail to prioritize high frequency functions effectively since there is no defined analysis on the distribution of container sizes in applications. 

\begin{figure}[h]
    \centering
    \includegraphics[width=1\columnwidth]{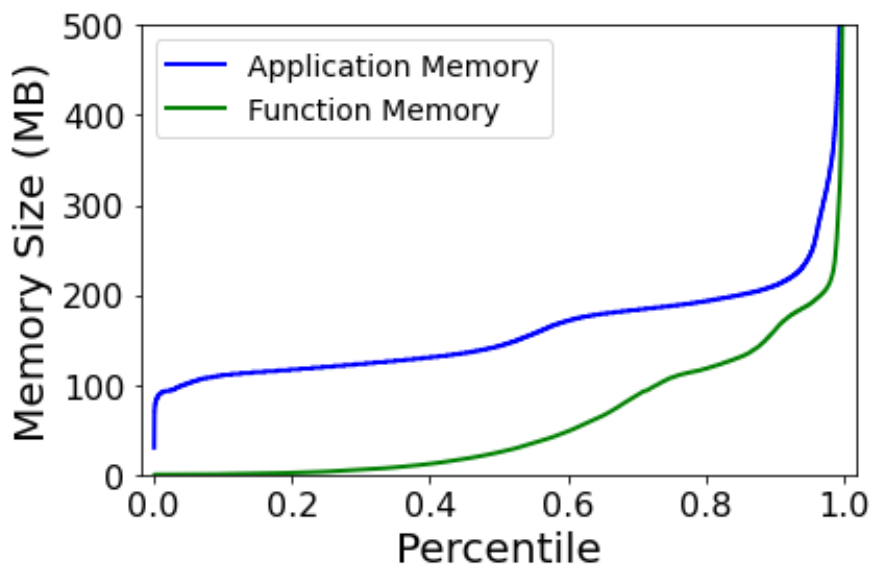}
    \caption{Percentile distribution of memory footprints for Azure Functions workloads.}
    \label{fig:memory_footprint}
\end{figure}

We plot the percentile distribution of application and function memory footprint as shown in Figure~\ref{fig:memory_footprint}.
We collect the application memory footprint from the Azure Functions data~\cite{mohammad_serverless_2020} and estimate the function memory using Equation~\ref{eq:func_mem}. 
The results show that more than 98\% of functions with small memory footprint consume below 225MB while large functions consume upto 500MB of memory.

Our analysis of Azure Functions data~\cite{mohammad_serverless_2020} identified a memory footprint spike at around 225 MB. This was determined using percentile distributions of application memory data across the 12 days available in the dataset, followed by the estimation of function memory:
\begin{equation}\label{eq:func_mem}
    \text{Function Memory} = \frac{\text{Application Memory} \times \text{Function Duration}}{\text{Application Duration}}
\end{equation}

\subsubsection{Traffic Frequency Correlation}
Frameworks such as \textit{Hermes}~\cite{kaffes_practical_2021} and \textit{HotC}~\cite{suo_tackling_2021} emphasize the importance of traffic-aware resource allocation.
However, they fail to account for the interplay between container size and invocation patterns. This leads to inefficiencies in caching policies.

\begin{figure}[h]
    \centering
    \includegraphics[width=1\columnwidth]{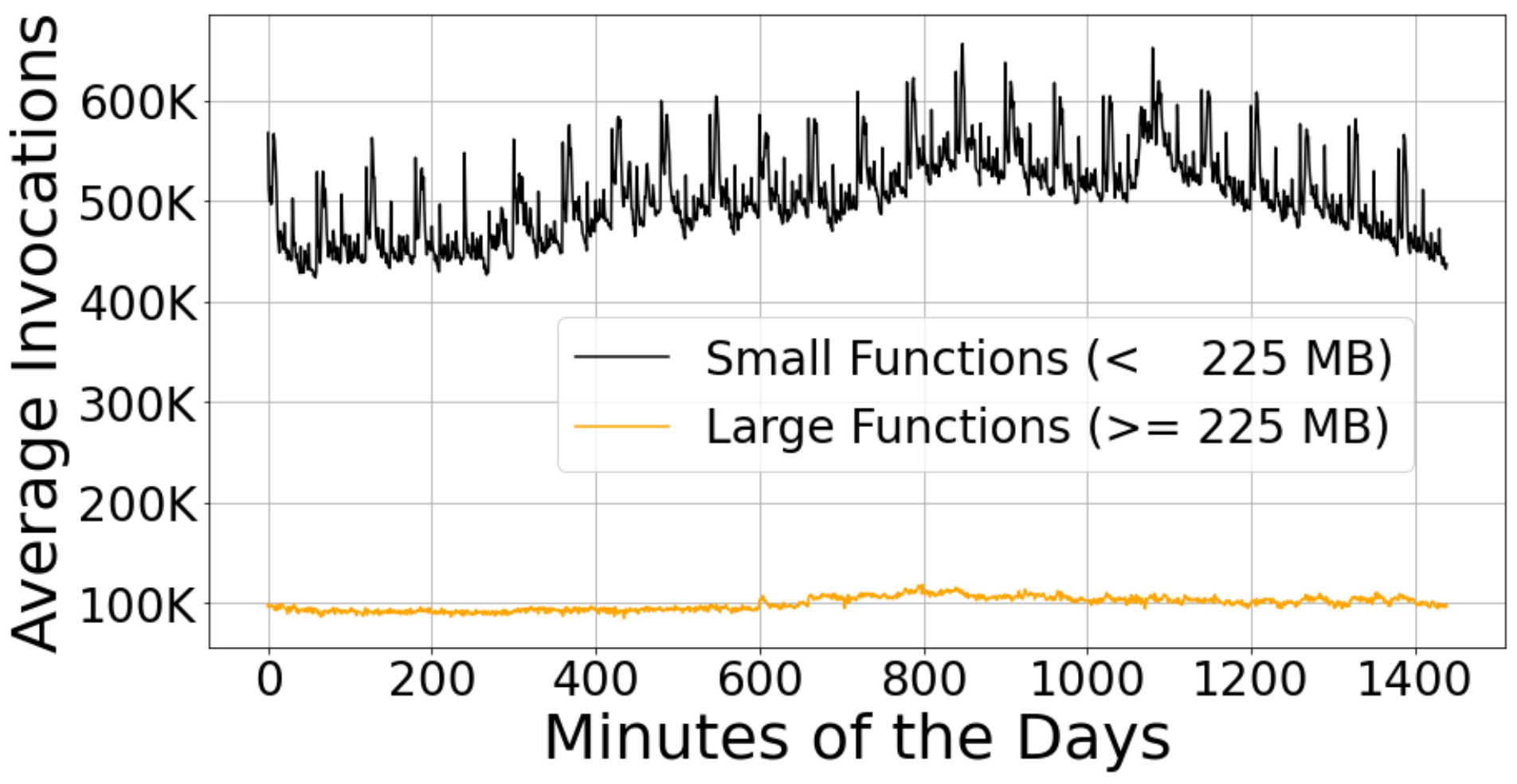}
    \caption{Normalized invocation trends for small and large functions.}
    \label{fig:traffic_frequency}
\end{figure}

Hence, we analyze the invocation frequency of small and large containers by categorizing the minute-by-minute invocation counts from the Azure Functions data with function memory sizes. 
Figure~\ref{fig:traffic_frequency} shows the invocation results over different times of a day.
The results show a clear distinction between the invocation frequency of small and large functions resulting in a 4–6.5$\times$ the number compared to large functions at any given time of the day. 

\subsubsection{Invocation Patterns and Inter-Arrival Times (IATs)}

We analyzed inter-arrival times (IATs) for Azure Functions using a sliding window approach. This method computed IATs within defined time windows (default: 60 minutes) with overlapping intervals (30 minutes) for smooth transitions. Outliers were filtered using a Z-score threshold to remove anomalies.

\begin{figure}[h]
   \centering
   \includegraphics[width=1\columnwidth]{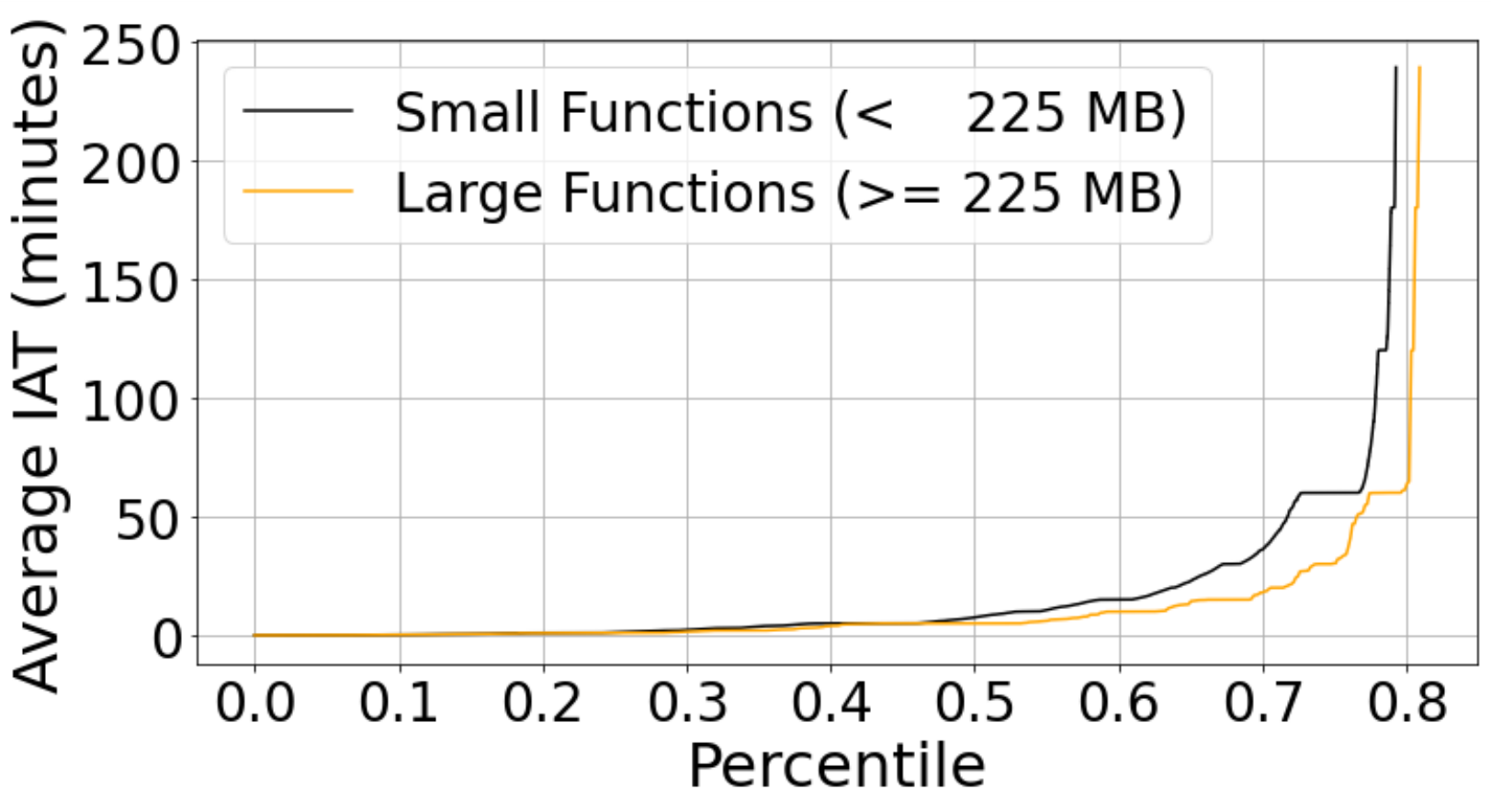}
   \caption{Percentile distribution of inter-arrival times for small and large functions.}
   \label{fig:iat_distribution}
\end{figure}

The average IAT distribution for small and large functions is shown in Figure~\ref{fig:iat_distribution}. The results reveal that:
\begin{itemize}
   \item Large functions invoke at similar or better intervals than small functions, especially at higher percentiles.
   \item Despite similar periodicity, the sheer volume of small functions leads to resource contention, exacerbating cold starts for large containers.
\end{itemize}

\subsubsection{Cold Start Latency and Resource Contention}
Prioritizing caching of small functions is essential because (i)~small functions experience cold start delays if not prioritized for caching, and (ii)~caching large functions instead of small functions results in excessive resource contention since these functions not only consume large amount of memory but also have longer runtimes. 

\begin{figure}[h] 
    \centering
    \includegraphics[width=1\columnwidth]{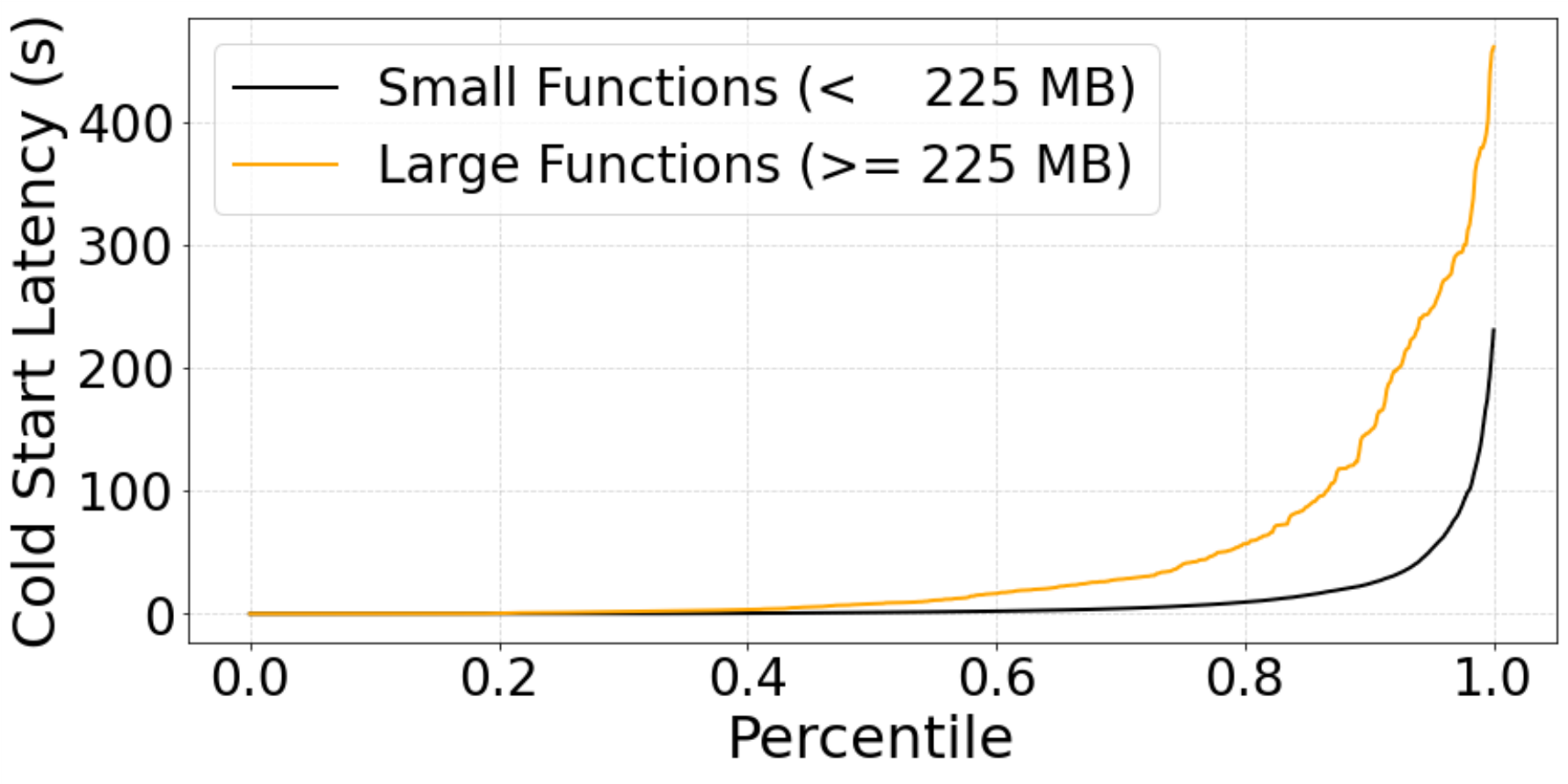}
    \caption{Percentile distribution of cold start latency for small and large functions.}
    \label{fig:cold_start_latency}
\end{figure}

However, it is essential to ensure proper caching of large functions also because large functions tend to have longer cold start delays. 
To demonstrate this, we analyze the cold start latency of large vs. small functions and plot the percentile distribution of latency in Figure~\ref{fig:cold_start_latency}. 
The distribution reveals that the large functions have longer latency with large functions exhibit latency of upto 100 seconds, compared to upto 15 seconds for small functions at the 85th percentile.

\section{Design}\label{sec:design}

%%%%%%%%%%%%%%%%%%%%%%%%%%%%%%

\begin{figure*}[t]
    \centering
    \includegraphics[trim = 15 530 15 15, width=1\textwidth]{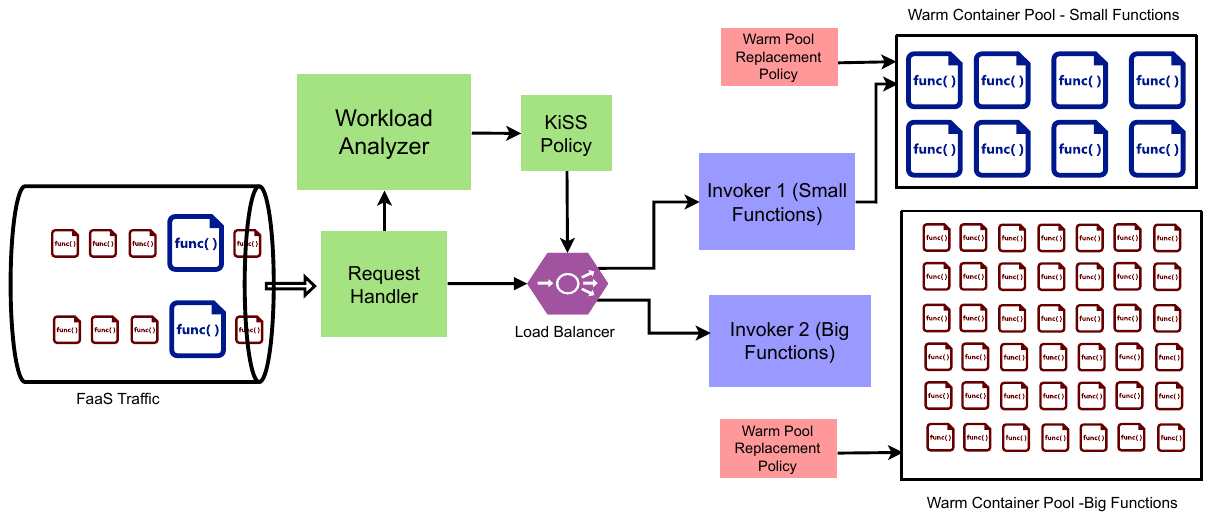}
    \caption{Overview of KiSS}
    \label{fig:overview}
\end{figure*}

The results we gleaned from the previous section (see Section~\ref{sec:work_anly}) helped in developing our policy: KiSS. The KiSS or \textbf{Keep it Separated Serverless} policy aims to address critical challenges in Function-as-a-Service (FaaS) platforms, particularly in edge computing environments, by achieving the following objectives:

\begin{itemize}
    \item \textbf{Reduced Cold Start Latency:} Prioritizes high-frequency functions to minimize delays in real-time applications.
    \item \textbf{Improved Resource Efficiency:} Optimizes memory and compute usage while avoiding unnecessary overhead from static warm states.
    \item \textbf{Minimized Inter-Function Interference:} Enhances throughput and scalability through modular resource partitioning.
    \item \textbf{Improved Function Service Rate:} Adopts resource-aware policies to reduce dropped requests and maximize system reliability.
\end{itemize}

\subsection{KiSS Policy Overview}

KiSS introduces a modular, data-driven orchestration strategy designed to optimize serverless execution in resource-constrained environments, particularly at the edge. By leveraging our workload analysis (refer Section 2.5), our policy segments functions based on key metrics—memory footprint, invocation frequency, and execution time—to optimize performance across diverse workloads.

The edge computing context introduces unique challenges like limited memory, heterogeneous resources, and dynamic workloads. Generalized cloud strategies often fail to adapt to such constraints. KiSS addresses this gap by analyzing workload characteristics and implementing a resource-efficient, modular strategy that aligns with edge-specific demands.

\subsection{Components of KiSS Policy Design}
Figure~\ref{fig:overview} shows the overall architecture of KiSS. 
The incoming \textit{FaaS traffic} will include both small and large functions. 
The \textit{request handler} accepts the incoming functions and shares the function information to the workload analyzer. 
The \textit{workload analyser} processes the function information to profile the incoming function traffic information and generate data such as invocation frequency, memory footprint etc.
The \textit{KiSS policy} uses this data to estimate where this function will be placed between the two different warm pool partitions.

The \textit{load balancer} implements a partitioning logic where functions are allocated to distinct warm pools using (\textit{invoker 1} and \textit{invoker 2}) based on profiling thresholds:

(i)~Small Functions Pool: Dedicated to high-frequency, low-memory functions to ensure low latency, and (ii)~Large Functions Pool: Allocated for low-frequency, memory-intensive functions, minimizing contention with smaller containers.
Each warm pool operates autonomously achieving Policy Independence.
The \textit{Warm Pool Replacement Policy} for each warm container pool can independently implement different workload-specific strategies to reduce contention and enhance temporal locality.

These factors form the foundation of KiSS’s multi-tiered warm pool framework, allowing it to effectively manage serverless resources and enhance performance in edge computing. By addressing these challenges, KiSS positions itself as a practical and scalable solution for FaaS platforms in environments with diverse and demanding resource constraints.

\subsection{Innovations of KiSS Policy}

One of the most innovative features of KiSS is its multi-level warm pool partitioning, which isolates high- and low-frequency functions into separate pools. This design eliminates inefficiencies inherent in monolithic resource strategies by ensuring that small, frequently invoked functions are always ready to execute, while larger, less frequent functions remain accessible without competing for resources. This adaptability extends to the ability to add more pools as workload patterns evolve, making KiSS a flexible and future-proof solution. Moreover, its modular architecture supports diverse deployment scenarios, from centralized clouds to resource-constrained edge environments. Integration with traffic-aware schedulers ensures that KiSS maintains scalability and responsiveness even under fluctuating workloads.

\subsubsection{Advantages of KiSS}

The advantages of KiSS are particularly pronounced in edge environments. By keeping frequently accessed containers in warm states, it drastically reduces cold start latency, which is critical for real-time applications such as IoT and AI analytics. Static warm pool partitioning, based on workload analysis, optimizes memory usage by eliminating unnecessary overhead, ensuring that resources are used efficiently even in environments with stringent memory constraints. This strategy not only enhances performance but also reduces operational costs by consolidating memory usage and minimizing cold starts. KiSS’s platform-agnostic design further enhances its versatility, enabling seamless deployment across various serverless frameworks.

\section{Methodology}\label{sec:method}

In this section, we outline the experimental setup, and the methodology employed by KiSS to evaluate cold start metrics.

\subsection{Simulation Environment}
We develop an enhanced and modified version of the \textbf{FaaSCache Simulator}~\cite{alexander_faascache_2021} to evaluate the KISS framework, tailoring it to address serverless resource management challenges in low resource constraint environments. This discrete event simulator models Function-as-a-Service (FaaS) systems as a dynamic warm pool, enabling analysis of cold start mitigation and resource allocation strategies. Key modifications support KiSS’s modular design and workload-specific approach.

The simulation is conducted across memory pool sizes ranging from 1 GB to 80 GB to capture diverse deployment scenarios, from low-resource edge nodes to well-provisioned cloud setups. For this study, results focus on the 1–24 GB range, as beyond that point resources are not heavily constrained.

Our approach provides a controlled and reproducible environment for testing KiSS across a broad range of workloads, enabling detailed insights into its performance and applicability.

The static 80-20 split in this evaluation serves as a representative configuration to assess the benefits of partitioning in a simulated environment.
Future studies could explore adaptive partitioning strategies that dynamically adjust memory allocation in response to changing workload demands.

\subsection{Workloads and Traffic Patterns}

Workload evaluation for the KiSS framework was based on trace derived from the 2019 Azure Function trace dataset~\cite{mohammad_serverless_2020}. This dataset provides detailed traffic patterns for serverless computing workloads required for this study. While no real-world edge-specific FaaS trace is publicly available, our assumption is that core properties of FaaS workloads remain consistent when extended to edge environments. Leveraging this assumption, we adapted the Azure trace to reflect the unique characteristics of edge deployments.

Container memory sizes were adjusted to align with typical edge constraints, with small containers ranging between 30--60 MB and large containers between 300--400 MB. The overall memory pool sizes were constrained to a range of 1--24 GB to reflect edge-specific hardware limitations. Invocation patterns and packet sizes were adapted to represent edge environments, focusing on smaller, frequent invocations (e.g., IoT event streams) alongside less frequent, resource-intensive tasks (e.g., video analytics).

\paragraph{Workload Diversity}
The synthesized trace allowed for the evaluation of KiSS under diverse conditions, including:
\begin{itemize}
    \item \textbf{High-Frequency Functions:} Representing lightweight, frequently invoked tasks such as IoT data processing.
    \item \textbf{Low-Frequency Functions:} Reflecting resource-intensive workloads like batch processing or analytics.
    \item \textbf{Bursty Traffic Patterns:} Simulating real-world traffic spikes, critical for understanding the framework’s behavior under sudden load surges.
    \item \textbf{Steady-State Operations:} Providing a baseline for performance under consistent invocation patterns.
\end{itemize}

By adapting the Azure trace to mimic edge-specific features, this evaluation captures a realistic approximation of how FaaS platforms would operate in constrained edge environments. This methodology bridges the gap between existing public cloud traces and the unique characteristics of edge deployments, ensuring that the insights gained are relevant to real-world applications.

\subsection{Evaluation Metrics}

The KiSS framework’s performance was evaluated using the following metrics:
\begin{itemize}
    \item \textbf{Cold Start Percentage:} The proportion of invocations requiring container initialization, critical for latency-sensitive applications.
    \item \textbf{Drop Percentage:} The proportion of invocations dropped due to memory contention, analyzed separately for small and large containers.
      \item \textbf{Fairness:} Performance consistency across small and large containers, ensuring equitable resource allocation.
\end{itemize}

\subsection{Fairness Analysis}

Fairness in resource allocation is a critical consideration for KiSS, as it ensures that both high-frequency (small) and low-frequency (large) containers receive equitable access to resources. Without careful planning, partitioned memory pools could lead to imbalances where one category of functions dominates resources. To address this, we conducted a fairness (equity) analysis to validate the effectiveness of the KiSS design in maintaining balance and meeting diverse service requirements.

The analysis focused on three key aspects. First, we evaluated equity in resource distribution to ensure that both small and large containers achieve comparable performance improvements. Second, we examined the avoidance of resource monopolization, ensuring that high-frequency small containers do not dominate memory resources at the expense of resource-intensive workloads. Lastly, we assessed support for diverse Quality of Service (QoS) requirements, validating that both latency-sensitive and resource-heavy functions could meet their respective performance goals.

This fairness analysis was performed by comparing cold start percentages and drop percentages across small and large containers. By doing so, we aimed to ensure that the KiSS framework delivers consistent and balanced performance across all workload categories, aligning with our goal of optimizing serverless execution in diverse and resource-constrained environments.

Fairness was assessed by comparing cold start percentages, drop percentages, and Policy performance across small and large containers.

\subsection{Policy Evaluation and Baseline Comparison}

The KiSS framework was evaluated with three different caching policies, alongside a unified warm pool as the baseline configuration. These evaluations aimed to validate the modularity and adaptability of KiSS across diverse resource management strategies:
\begin{itemize}
    \item \textbf{Least Recently Used (LRU):} The default policy, applied uniformly in the baseline and within partitioned memory pools for KiSS~\cite{jonas_cloud_2019,alexander_faascache_2021}.
    \item \textbf{Greedy Dual(GD):} A Greedy Dual policy inspired by FaaSCache~\cite{alexander_faascache_2021} that incorporates multiple features like invocation frequency and memory footprint to make eviction decisions.
    \item \textbf{Frequency-Based (Freq):} A policy that prioritizes caching for frequently invoked functions, irrespective of resource type~\cite{alexander_faascache_2021}.
\end{itemize}

The baseline configuration used a unified warm pool with the LRU caching policy, treating all containers equally. KiSS was tested with the same policies (LRU, GD, and Freq) to measure the relative benefits of its partitioned architecture.

\section{Implementation}\label{sec:implement}

The effectiveness of the KiSS policy relies on fine-tuning three critical parameters: container size thresholds, memory size per pool, and total memory allocation, explained in the following section. 

\subsection{Tuning Parameters}

Parameters were systematically refined within a simulation environment to optimize performance, minimize resource contention, and effectively address cold start latency.

\subsubsection{Container Size Thresholds}
The categorization of containers into \textit{small} and \textit{large} based on memory footprint was central to our policy's design. The calibration process for container size thresholds considered our central paradigm of keeping functions separated (low and high frequency). The calibration process involved:

\begin{itemize}
    \item \textbf{Empirical Benchmarking:} Initial thresholds were derived from the distribution of function memory footprints, aligning with observed workload patterns.
    \item \textbf{Simulation Validation:} Various threshold values were tested to evaluate their impact on latency, memory usage, and throughput.
    \item \textbf{Dynamic Adjustments:} Iterative simulations ensured that small containers achieved low-latency performance while large containers benefited from efficient inter-function interference handling mechanism.
\end{itemize}

\subsubsection{Memory Size Per Pool}
Allocating appropriate memory resources for each warm pool was critical to balancing readiness and efficiency:
\begin{itemize}
    \item \textbf{Small Container Pool:} This pool was allocated a larger cache share to accommodate high-frequency, latency-sensitive functions, with dynamic adjustments made to handle traffic spikes and ensure low-latency performance.
    \item \textbf{Large Container Pool:} A smaller cache allocation was designated for this pool, relying on snapshot-based provisioning to manage sporadic invocations without consuming excess memory resources.
\end{itemize}

The tuning process for cache size included:
\begin{itemize}
    \item \textbf{Hit Rate Optimization:} The simulator tracked container warm pool hit rates for each pool under varying workloads to identify the optimal allocation that minimized cold start occurrences.
    \item \textbf{Latency Analysis:} Observing response times helped fine-tune memory-footprint distributions, ensuring resource efficiency without compromising performance.
\end{itemize}

\subsubsection{Total Memory Size}
The overall memory pool capacity was calibrated to ensure function readiness without over-provisioning. This helps us to study different resource environments, like Edge:
\begin{itemize}
    \item \textbf{Stress Testing:} Workload bursts traces were tested memory size adequacy under varying demand.
    \item \textbf{Iterative Refinement:} Adjustments balanced latency reduction and memory utilization, ensuring consistent performance across diverse workloads.
\end{itemize}

\subsection{Metrics for Performance Evaluation}
The simulator tracks six key metrics to quantify performance across diverse configurations:
\begin{enumerate}
    \item \textbf{Cold Starts (Misses):} Instances where a new container must be initialized because no matching container exists in the resource pool.
    \item \textbf{Hits:}  Function invocations that successfully utilize an existing container, avoiding a cold start.
    \item \textbf{Drops:} Scenarios where a missed function cannot allocate a new container due to all containers being actively utilized. This extended metric offers deeper insights into resource contention.
    \item \textbf{Total Accesses:} The total number of function invocations, encompassing hits, misses, and drops.
    \item \textbf{Serviceable Accesses:} Invocations that were successfully serviced, combining hits and misses.
    \item \textbf{Execution Durations:} The cumulative execution time for all functions, calculated from cold start and warm start durations.
\end{enumerate}

These metrics provide a comprehensive framework for evaluating resource allocation efficiency and enable rigorous comparisons with baseline setups and existing state-of-the-art methods.

\section{Results}\label{sec:results}

\subsection{Cold Start Percentage}

\begin{figure}[h]
    \centering
    \includegraphics[trim = 7 5 5 5, width=1\columnwidth]{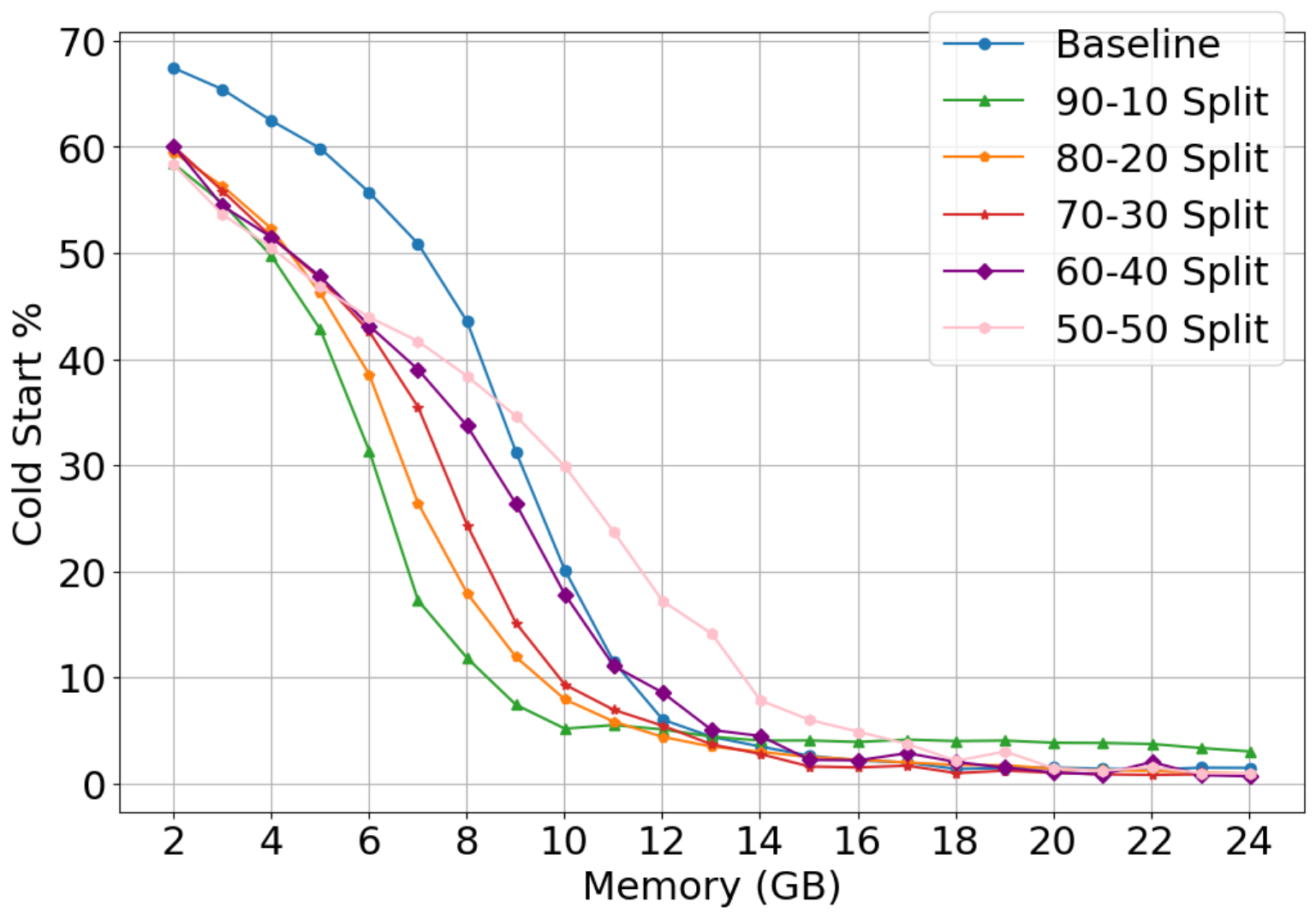}
    \caption{Cold start percentages across different configurations.}
    \label{fig:cold_start_1}
\end{figure}

\begin{figure}[h]
    \centering
    \includegraphics[width=1\columnwidth]{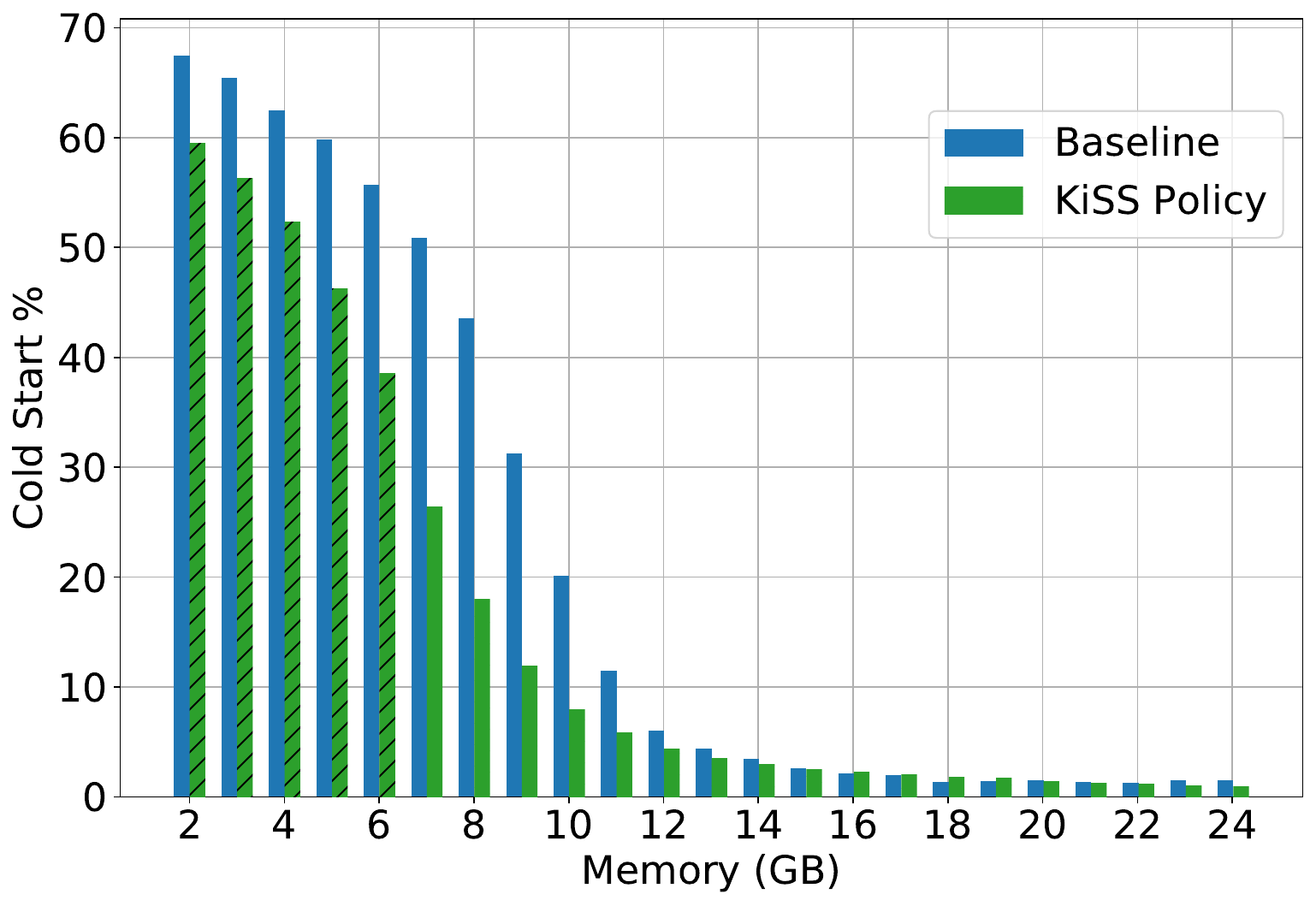}
    \caption{Comparison of the 80-20 split with the baseline configuration.}
    \label{fig:cold_start_2}
\end{figure}

Figures~\ref{fig:cold_start_1} and~\ref{fig:cold_start_2} illustrate the impact of the KiSS framework's partitioned warm pool architecture on cold start percentages under various configurations, including 90-10, 80-20, 70-30, 60-40, and 50-50 splits, compared to a baseline without partitioning. The 80-20 split consistently achieved the lowest cold start percentages, especially in memory-constrained edge environments (4–16 GB):
\begin{itemize}
    \item \textbf{4 GB memory:} Cold starts dropped from \textbf{62\% (baseline)} to \textbf{52\%}, a \textbf{16.2\% improvement}.
    \item \textbf{8 GB memory:} Cold starts decreased from \textbf{43\% to 18\%}, a \textbf{58\% reduction}.
    \item \textbf{10 GB memory:} Cold starts were reduced from \textbf{20\% to 8\%}, a \textbf{60\% improvement}.
\end{itemize}

Alternative splits showed varying degrees of performance. The 70-30 split, while close in performance, exhibited higher cold start percentages in low-memory settings (e.g., \textbf{42\% at 4 GB} vs. \textbf{38\% for 80-20}). The 90-10 split overly prioritized small containers, starving large containers of resources, while the 50-50 split failed to sufficiently prioritize small containers, resulting in significantly higher cold start percentages overall.

The scalability of the 80-20 split was particularly notable. In highly provisioned environments (>16 GB), cold start percentages for both the 80-20 and baseline configurations approached near-zero, showing diminishing returns in memory-abundant scenarios.

\subsection{Drop Percentage}

\begin{figure}[h]
    \centering
    \includegraphics[width=1\columnwidth]{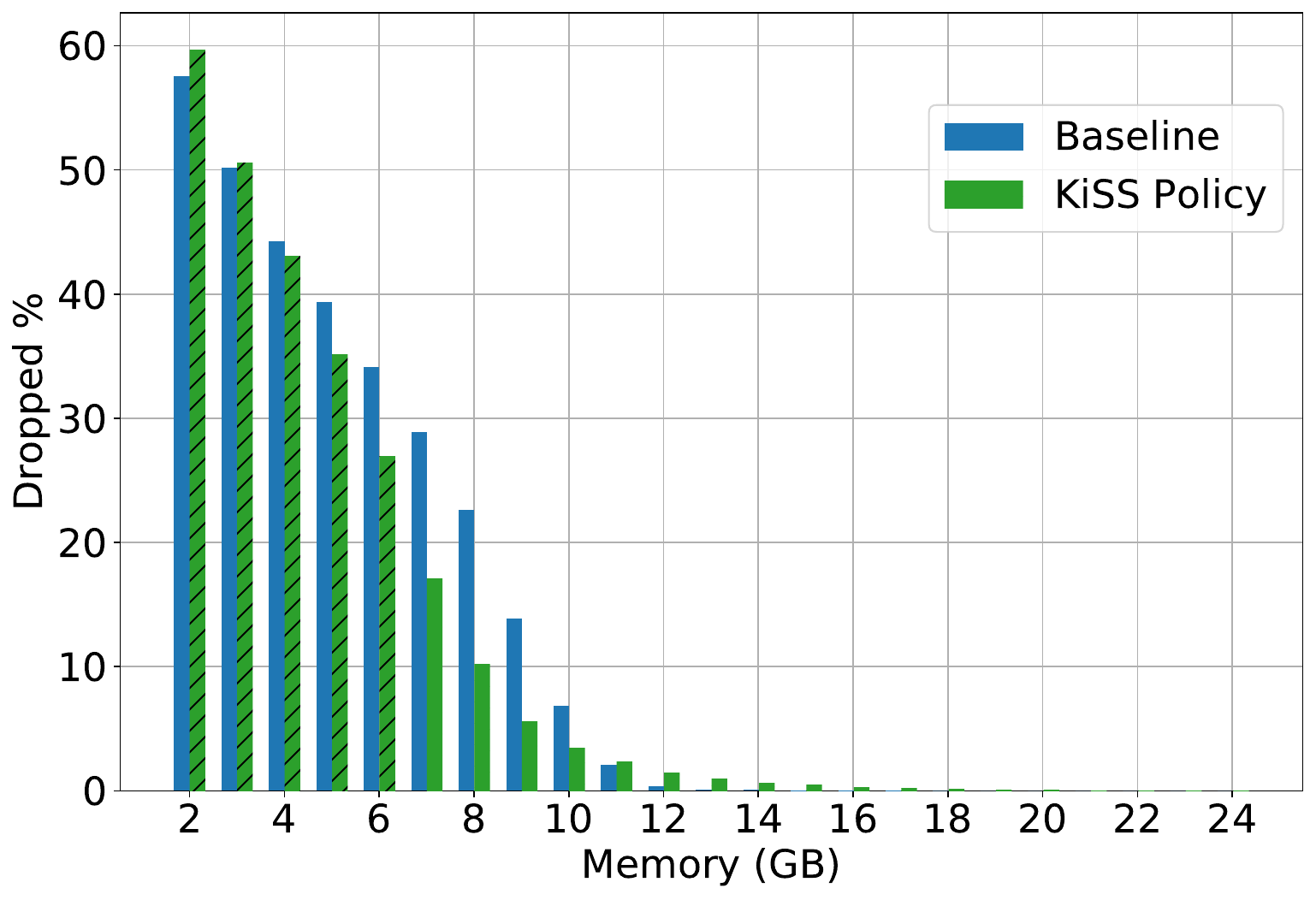}
    \caption{Drop percentage across memory configurations.}
    \label{fig:drop_percentage}
\end{figure}

Figure~\ref{fig:drop_percentage} highlights the drop percentage, which measures the proportion of function invocations that cannot be serviced due to resource contention.
\begin{itemize}
    \item \textbf{2–3 GB memory:} KiSS showed slightly higher drop percentages than the baseline, with drops at \textbf{60\% vs. 58\%} for 2 GB and \textbf{51\% vs. 50\%} for 3 GB. This was attributed to the resource isolation introduced by partitioning in extremely low-memory settings.
    \item \textbf{4–8 GB memory:} Partitioning stabilized, and KiSS significantly reduced drops. At \textbf{6 GB}, drops decreased from \textbf{34\% (baseline)} to \textbf{27\%} (\textbf{21\% improvement}). At \textbf{8 GB}, drops fell from \textbf{23\% to 10\%}, reflecting a \textbf{56.5\% improvement}.
    \item \textbf{Beyond 8 GB:} Both configurations exhibited near-zero drop percentages, demonstrating the scalability of the KiSS framework in well-provisioned environments.
\end{itemize}

\subsection{Fairness Analysis}

\begin{figure}[h]
    \centering
    \includegraphics[width=1\columnwidth]{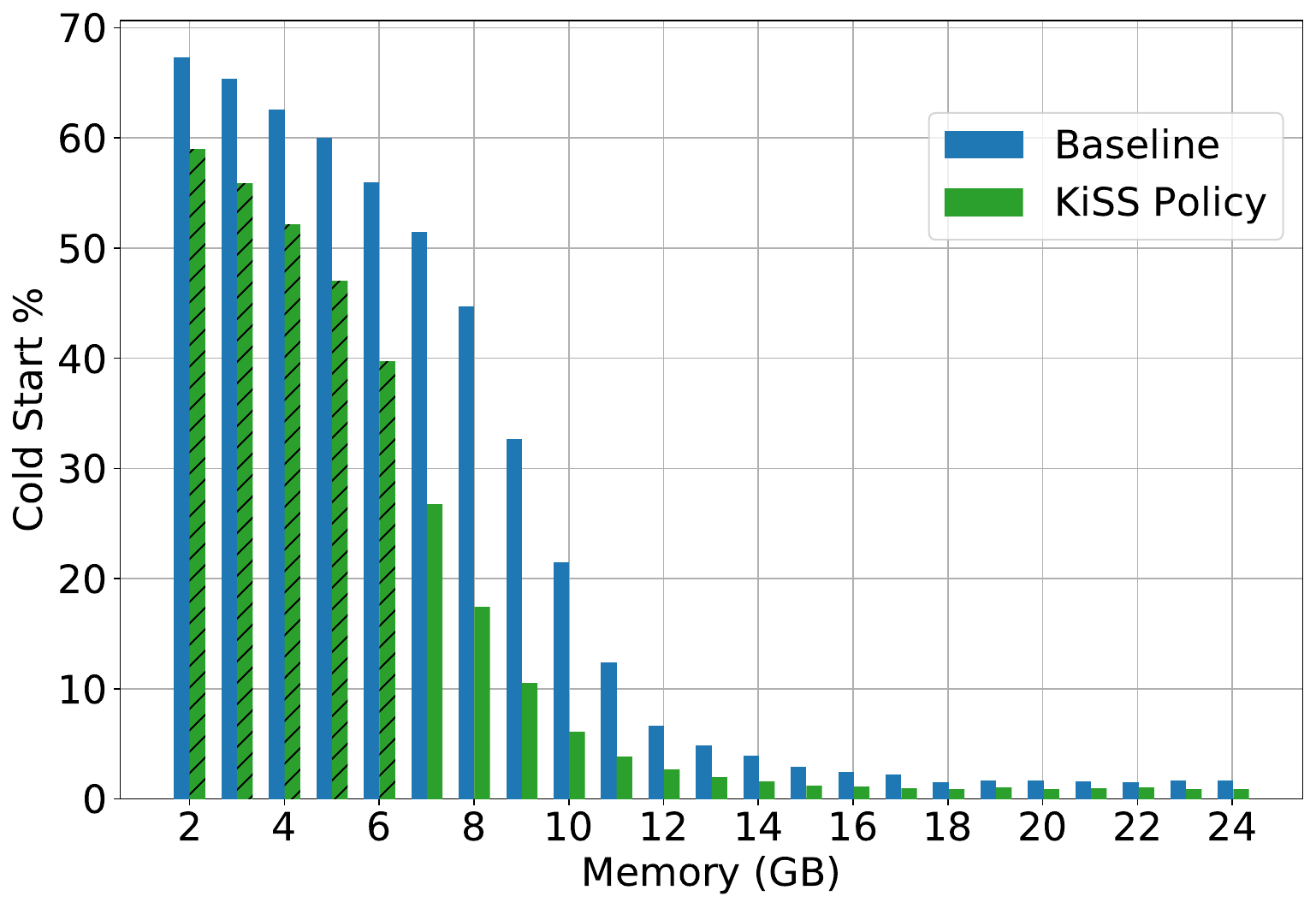}
    \caption{Cold start percentages for small containers.}
    \label{fig:cold_start_small}
\end{figure}

\begin{figure}[h]
    \centering
    \includegraphics[width=1\columnwidth]{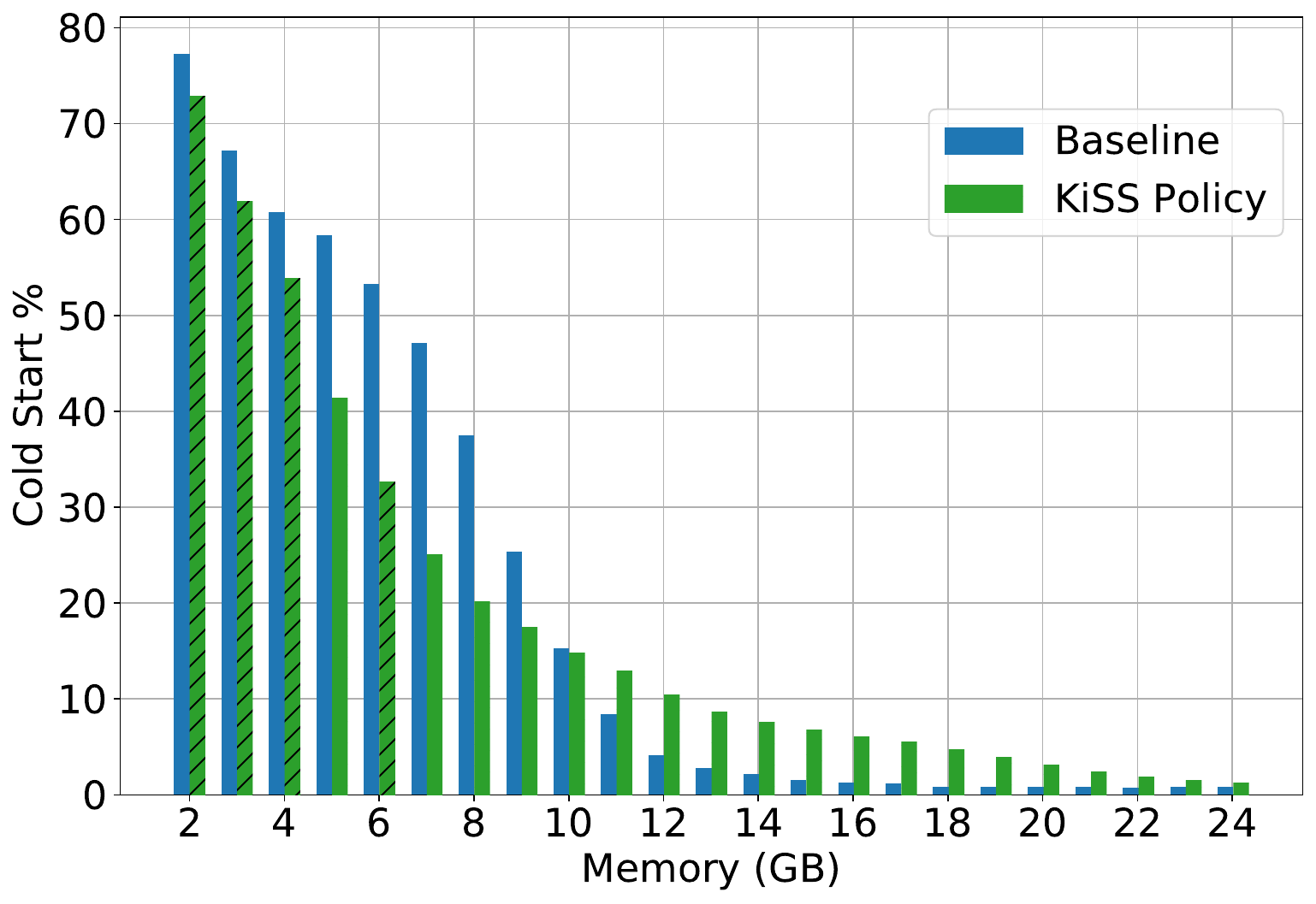}
    \caption{Cold start percentages for large containers.}
    \label{fig:cold_start_large}
\end{figure}

\begin{figure}[h]
    \centering
    \includegraphics[width=1\columnwidth]{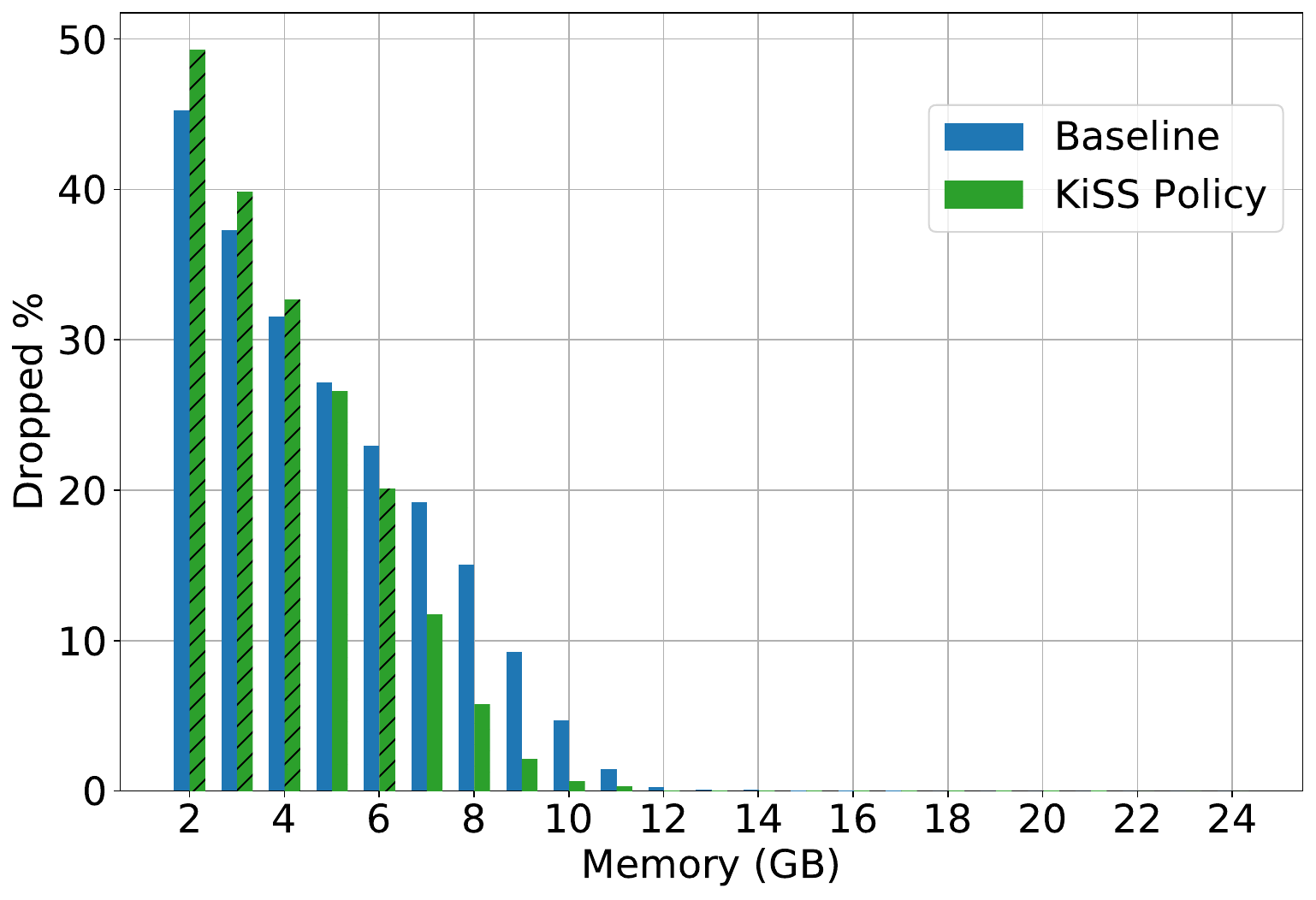}
    \caption{Drop percentages for small containers.}
    \label{fig:drop_small}
\end{figure}

\begin{figure}[h]
    \centering
    \includegraphics[width=1\columnwidth]{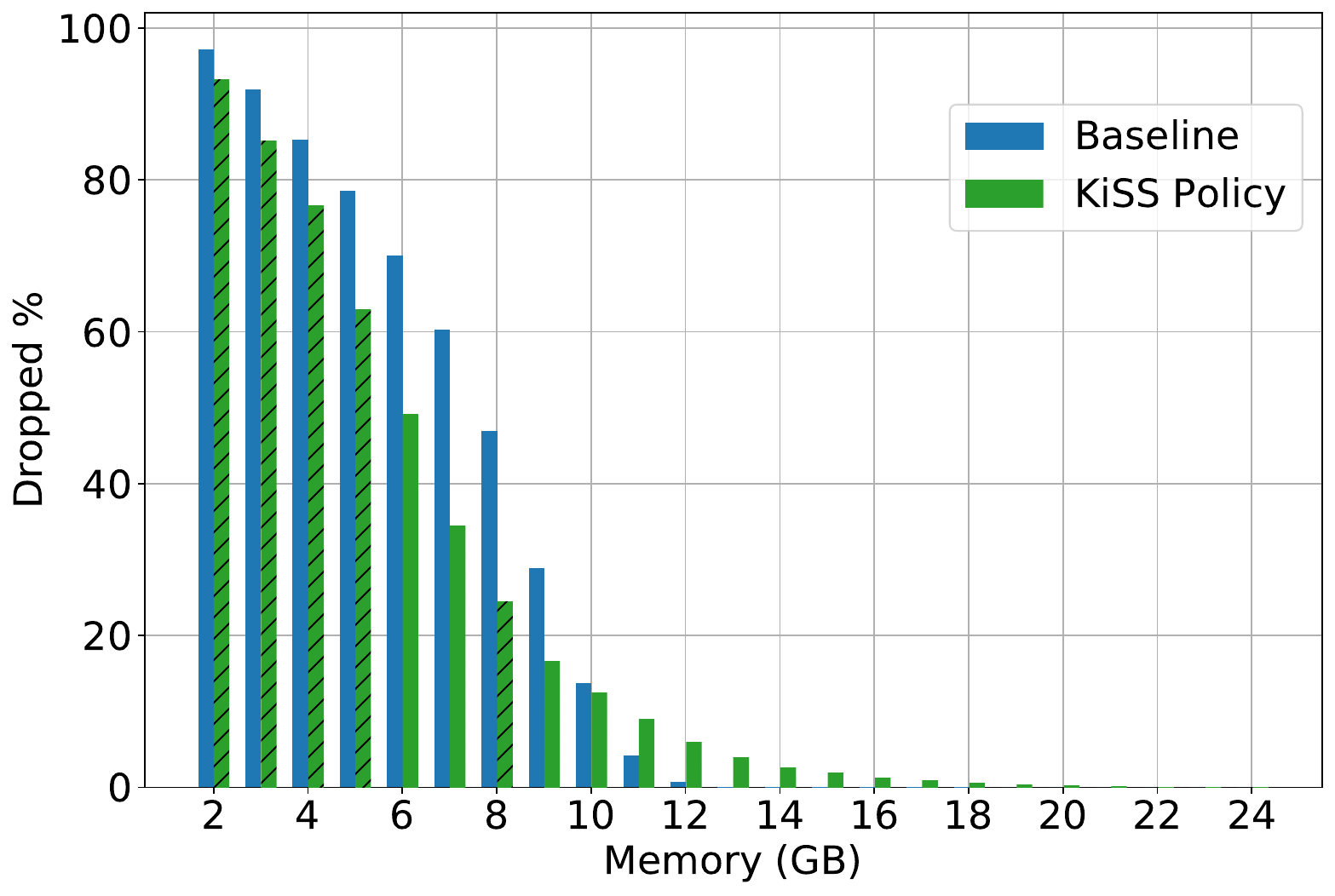}
    \caption{Drop percentages for large containers.}
    \label{fig:drop_large}
\end{figure}

Figures~\ref{fig:cold_start_small} through~\ref{fig:drop_large} assess fairness by comparing cold start percentages, drop percentages, and memory utilization for small (QoS) and large (QoSLarge) containers.

Figures~\ref{fig:cold_start_small} and~\ref{fig:cold_start_large} illustrate the following trends for cold start percentages:
    \begin{itemize}
        \item \textbf{Small Containers:} At \textbf{4 GB}, cold starts reduced from \textbf{63\% (baseline)} to \textbf{53\%}, a \textbf{16\% improvement}, and from \textbf{45\% to 18\%} at \textbf{8 GB}, a \textbf{60\% reduction}.
        \item \textbf{Large Containers:} At \textbf{4 GB}, cold starts dropped from \textbf{61\% to 54\%} (\textbf{11.5\% improvement}) and from \textbf{37\% to 20\%} at \textbf{8 GB} (\textbf{46\% improvement}).
    \end{itemize}

Figures~\ref{fig:drop_small} and~\ref{fig:drop_large} illustrate the following trends in drop percentages for small and large containers:
    
    \begin{itemize}
        \item \textbf{Small Containers:} Drops increased slightly at \textbf{4 GB} (\textbf{32\% to 33\%}) but improved significantly at \textbf{8 GB} (\textbf{15\% to 6\%}, a \textbf{60\% improvement}).
        \item \textbf{Large Containers:} Drops reduced from \textbf{85\% to 78\%} at \textbf{4 GB} (\textbf{8.2\% improvement}) and from \textbf{47\% to 24\%} at \textbf{8 GB} (\textbf{49\% improvement}).
    \end{itemize}

\subsection{Policy Independence}

\begin{figure}[h]
    \centering
    \includegraphics[width=1\columnwidth]{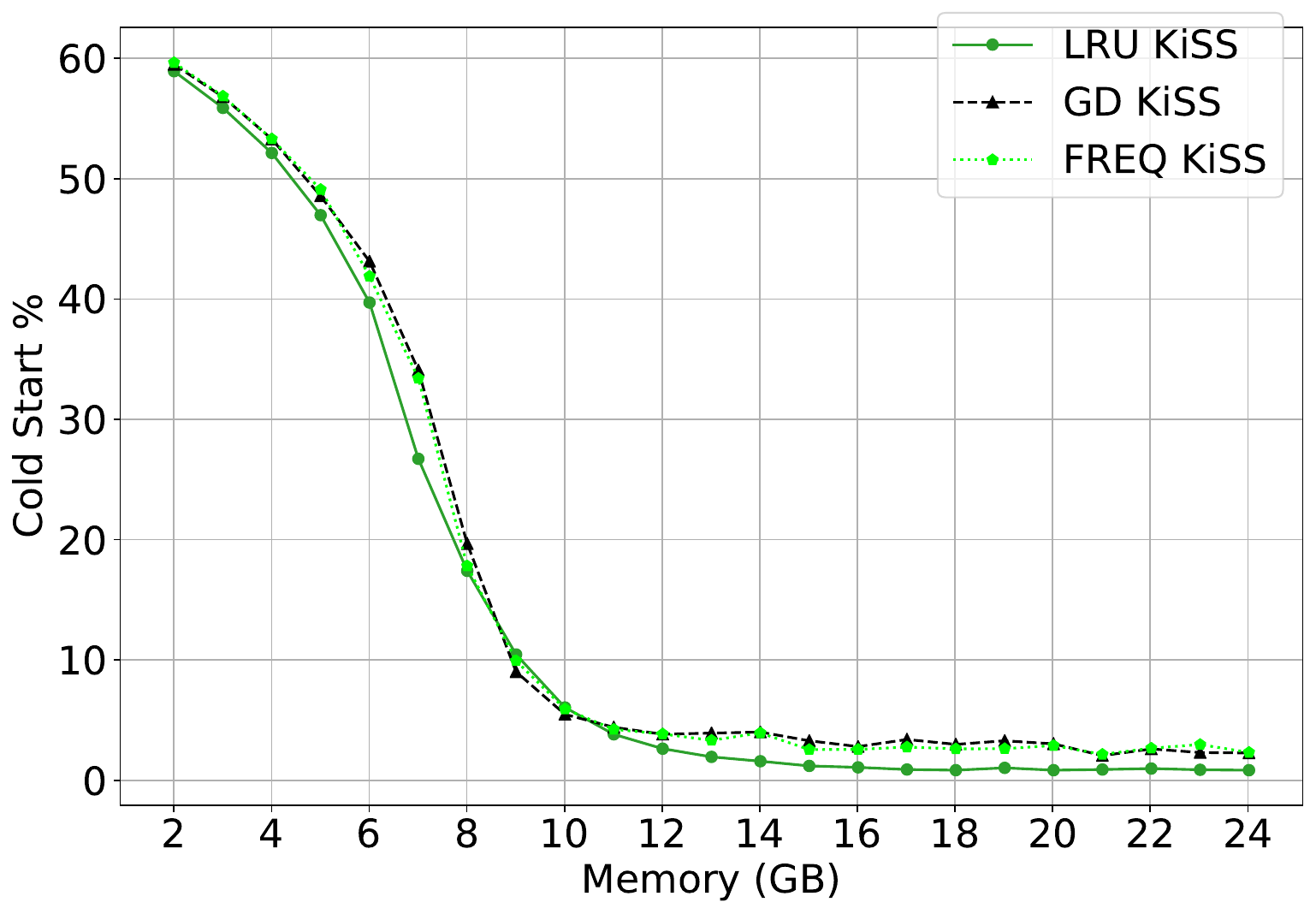}
    \caption{Cold start percentages for small containers across LRU, GD, and FREQ policies.}
    \label{fig:policy_fairness_small}
\end{figure}

\begin{figure}[h]
    \centering
    \includegraphics[trim = 5 35 20 30, width=1\columnwidth]{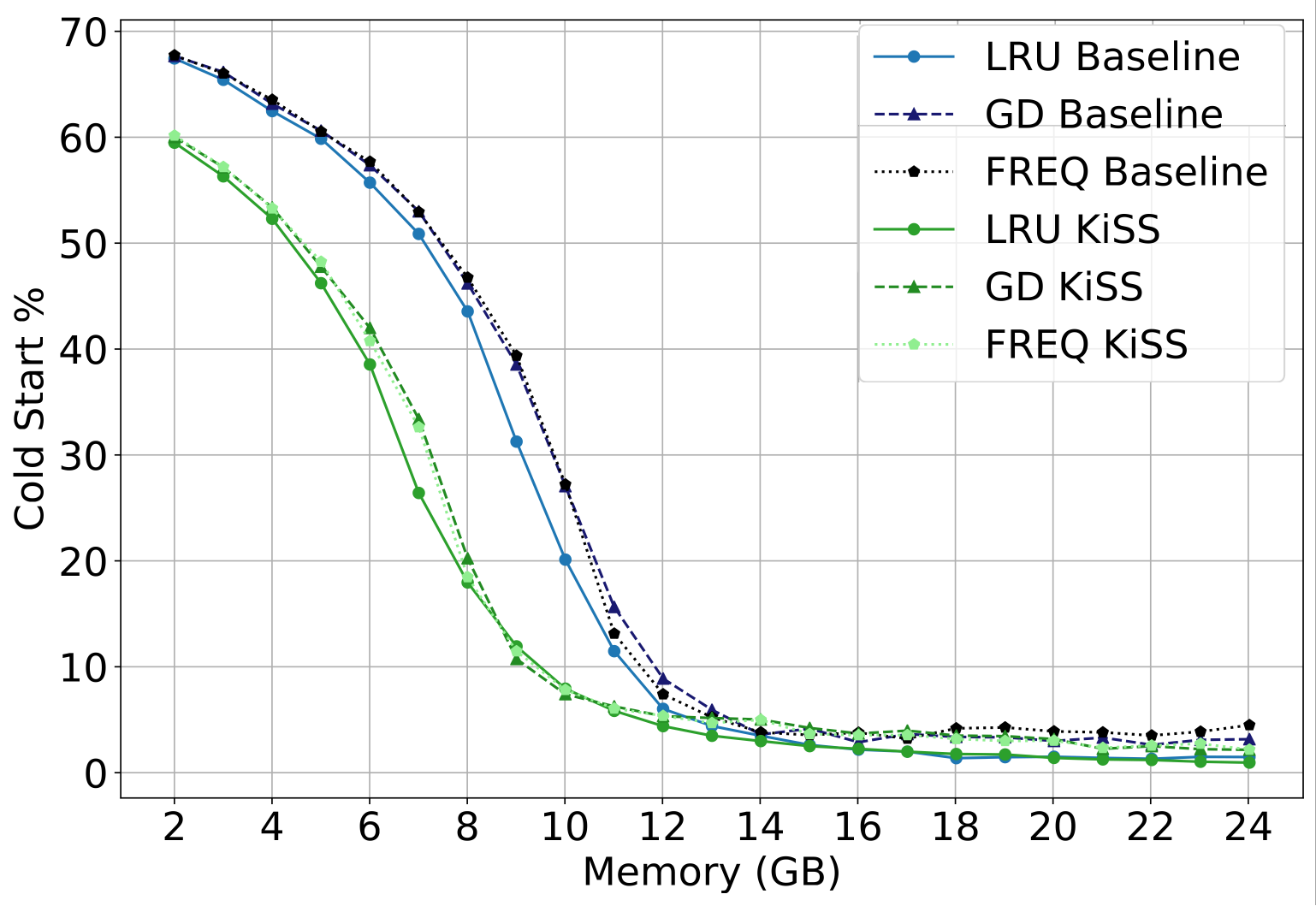}
    \caption{Overall cold start percentages across LRU, GD, and FREQ policies.}
    \label{fig:policy_overall}
\end{figure}

\begin{figure}[h]
    \centering
    \includegraphics[width=1\columnwidth]{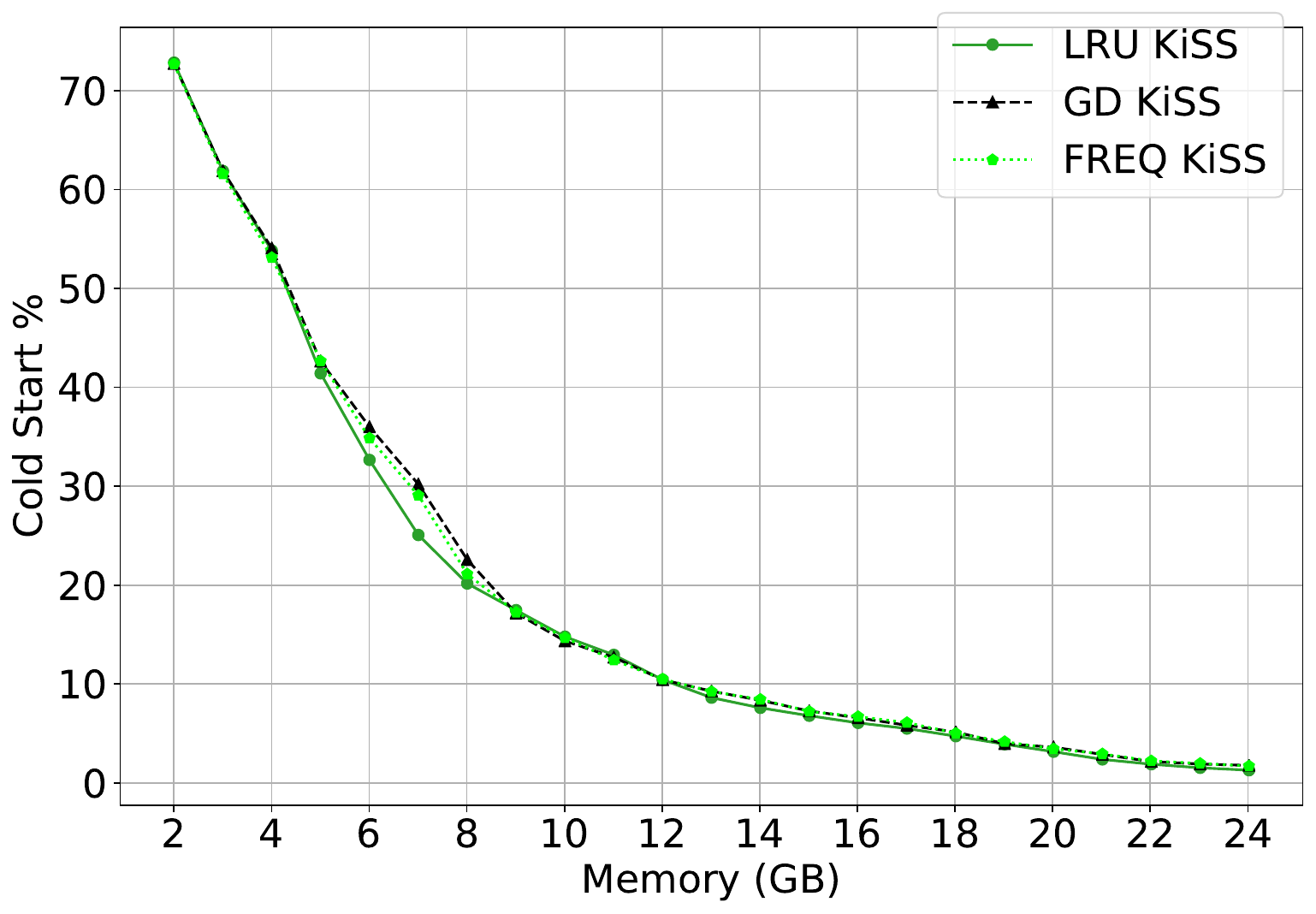}
    \caption{Cold start percentages for large containers across LRU, GD, and FREQ policies.}
    \label{fig:policy_fairness_large}
\end{figure}

The KiSS framework demonstrates robust independence from specific replacement policies, maintaining consistent performance across Least Recently Used (LRU), Greedy Dual (GD), and Frequency-Based (FREQ) policies. This flexibility makes KiSS adaptable to a variety of serverless environments, allowing it to optimize performance regardless of the resource management strategy in use. Figures~\ref{fig:policy_fairness_small}, \ref{fig:policy_overall}, and \ref{fig:policy_fairness_large} illustrate the framework’s consistent cold start percentage reduction for small, overall, and large containers, respectively.

\noindent\textbf{Small Containers Performance}  
Figure~\ref{fig:policy_fairness_small} shows the cold start percentages for small containers under the three replacement policies. Across the memory configurations, all policies exhibit similar trends. The cold start percentages reduce significantly as memory increases from 4 GB to 10 GB, with negligible differences between policies. In edge environments (4–8 GB memory), LRU slightly outperforms GD and FREQ, but the differences are marginal. This demonstrates KiSS’s ability to prioritize high-frequency small containers, regardless of the policy in use.

\noindent\textbf{Overall Performance}  
As shown in Figure~\ref{fig:policy_overall}, the overall cold start percentages for all containers (small and large) remain consistent across the three policies. LRU, GD, and FREQ exhibit overlapping performance trends, with cold start percentages converging to near-zero beyond 16 GB memory. In edge-specific memory ranges (4–8 GB), KiSS achieves significant cold start reductions under all policies when compared to baseline, emphasizing its independence from specific replacement strategies.

\noindent\textbf{Large Containers Performance}  
Figure~\ref{fig:policy_fairness_large} highlights the cold start percentages for large containers across the three replacement policies. In memory-constrained settings (4–6 GB), the differences between policies are slightly more pronounced, with GD and FREQ slightly underperforming compared to LRU. However, the differences diminish as memory scales, with all policies converging at near-zero cold start percentages beyond 16 GB. This shows that KiSS ensures adequate prioritization of low-frequency, high-memory functions, regardless of the policy used.

\subsection{Stress Testing}
To evaluate the robustness of KiSS under high-demand conditions, we conducted a stress test using a two-hour unedited trace comprising 4--5 million invocations on a 10 GB memory pool.  
KiSS serviced 150,000 requests compared to 160,000 in the baseline, maintaining high throughput under extreme load. KiSS also improved the hit rate from 0.38\% in the baseline to 2.85\%, showcasing its ability to prioritize critical requests and reduce contention.

Our analysis showed that the increased hit rate under KiSS demonstrates its ability to manage resource contention effectively. Additionally, KiSS maintained robust performance during workload spikes, validating its scalability and adaptability.

\section{Discussions and Conclusions}\label{sec:discuss}
In edge environments (4--8 GB), KiSS effectively reduced cold start percentages and drops compared to the baseline. Small containers showed cold start reductions of up to 30\%, while large containers achieved a 33\% improvement. Similarly, drops were nearly halved for small containers (47\%) and reduced by 40\% for large containers at 8 GB. These gains stem from workload-aware partitioning, which isolates resource pools to prioritize small, high-frequency containers.

While the static 80-20 split performed well, slight increases in drops were observed in very low memory ranges (2--3 GB) due to partitioning constraints. This trade-off suggests that adaptive partitioning could further optimize resource allocation under extreme constraints.

\subsection{Fairness and Resource Efficiency}

KiSS achieved equitable improvements for both small and large containers. Speedup ratios remained consistent across memory configurations, reflecting balanced resource utilization without overloading or underutilizing memory. This is particularly important for edge environments, where non-server-grade hardware requires careful resource management to avoid risks like thermal throttling and hardware wear.

\subsection{Implications for Edge FaaS Deployments}

The findings position KiSS as a robust solution for edge environments. Its ability to reduce cold starts, request drops, balancing fairness and preventing over-utilization, makes it particularly suitable for latency-sensitive applications such as IoT event processing and real-time analytics.

\subsection{Opportunities for Further Optimization}

While KiSS performed strongly, its reliance on static partitioning presents opportunities for improvement. Adaptive partitioning informed by real-time workload monitoring could address the observed trade-offs in very low memory ranges. Additionally, testing KiSS under highly variable and bursty traffic patterns would provide a broader perspective on its applicability. Incorporating reinforcement learning-based caching strategies could also enhance its ability to adapt dynamically to workload shifts.

The static 80-20 split in this evaluation serves as a representative configuration to assess the benefits of partitioning in a simulated environment. Future studies could explore adaptive partitioning strategies that dynamically adjust memory allocation in response to changing workload demands.

% \input{conclusion}

% This section has a special environment:
% \begin{verbatim}
%   \begin{acks}
%   ...
%   \end{acks}
% \end{verbatim}
% so that the information contained therein can be more easily collected
% during the article metadata extraction phase, and to ensure
% consistency in the spelling of the section heading.

% Authors should not prepare this section as a numbered or unnumbered {\verb|\section|}; please use the ``{\verb|acks|}'' environment.

%\input{appendix}

\bibliographystyle{ACM-Reference-Format}
\bibliography{main}

\end{document}